\documentclass[twocolumn,aps,superscriptaddress,multicol,amsmath,amssymb]{revtex4-2}

\usepackage{graphicx}
\usepackage{dcolumn}
\usepackage{bm}
\usepackage[colorlinks,linkcolor=blue,citecolor=blue]{hyperref}

\graphicspath{{figs/}}					

\begin{document}

\preprint{APS/123-QED}

\title{Topology of the Visibility Graph of Sandpiles}

\author{V. Adami}
\affiliation{Department of Physics, University of Mohaghegh Ardabili, P.O. Box 179, Ardabil, Iran}

\author{H. Masoomy}
\affiliation{Department of Physics, Shahid Beheshti University, 1983969411, Tehran, Iran}

\author{M. N. Najafi}
\email{morteza.nattagh@gmail.com}
\affiliation{Department of Physics, University of Mohaghegh Ardabili, P.O. Box 179, Ardabil, Iran}

\date{\today}

\begin{abstract}
In this paper, we explore the higher- and lower-order connectivity aspects of the visibility graph representation of sandpile models, focusing on the Bak-Tang-Wiesenfeld (BTW) model. By converting the time series of avalanche events into visibility graphs, we investigate the structural properties that emerge at different scales of connectivity. Specifically, we utilize persistent homology and simplicial complex theory to identify higher-order topological features. Analyzing the statistics of degree and betweenness centralities reveal that the resulting graph is a scale-free network with the degree exponent of $\gamma_k=2.50\pm 0.02$ and betweenness exponent of $\gamma_b=1.585\pm 0.008$. The analysis of the simplexes show that the distribution function of one, two and three dimensional simplexes are power-law with the exponents of $\gamma_{\sigma_1}=0.795\pm 0.006$, $\gamma_{\sigma_2}=0.602\pm 0.009$ and $\gamma_{\sigma_3}=0.422\pm 0.008$ respectively. This power-law behavior is also seen in the homology group analysis, for the Betti numbers, the exponents of which are reported in the paper. The persistent entropy of life time of $d$ dimensional holes is analyzed indicating that it increases logarithmically in terms of the network size $N$ for $d=0,1$ and $2$.
\end{abstract}

	
\maketitle

\section{Introduction}
Complex networks have long served as a powerful framework for analyzing dynamical systems. Examples include their application to earthquakes and seismic activities~\cite{abe2004small,abe2006complex,abe2007dynamical,abe2012dynamical}, chaotic time series~\cite{gao2012directed}, pseudoperiodic time series~\cite{zhang2008characterizing}, neuronal systems~\cite{sporns2011human,del2020network}, the topology of cortical connectivity in mammalian brains~\cite{zemanova2006structural}, mental disorders like stress~\cite{mcnally2015mental}, supply network theory~\cite{hearnshaw2013complex}, power grid infrastructures~\cite{pagani2013power}, transportation networks~\cite{xu2008exploring}, and clinical science~\cite{hofmann2016complex}. Through these applications, the statistical and topological characteristics of complex networks provide valuable insights into the hidden dynamics of these systems. Prominent tools in this field include synchronization~\cite{yu2009pinning,wu2007synchronization,song2010synchronization}, the eigenvalue spectrum of adjacency matrices, exponents in scale-free networks, homology and homotopy group structures, and various statistical measures~\cite{wang2016data}. Additionally, complex networks are instrumental in understanding time series structure of dynamical systems, yielding significant advancements~\cite{lacasa2008time}. This is achieved through the \textit{visibility graph} technique~\cite{lacasa2008time}, a robust method for transforming time series data into graph representations. In this approach, time points are treated as nodes, and edges are established based on geometric visibility criteria~\cite{lacasa2008time}, enabling effective analysis of dynamical systems. Examples of its application include seismic activities and earthquakes~\cite{telesca2012analysis}, geological time series~\cite{donner2012visibility}, Gaussian noise~\cite{masoomy2023relation}, wall turbulence~\cite{iacobello2018visibility}, macroeconomic time series in China~\cite{wang2012visibility}, and heartbeat dynamics~\cite{jiang2013visibility}. Recent studies have also demonstrated that the visibility graphs of sandpiles, fractional Brownian motion, and Lévy processes exhibit complex networks with rich structures. For a comprehensive review, see~\cite{turner2001isovists}.\\ 

The visibility graph transformation enables the application of graph-theoretical methods to explore the temporal correlations and structural patterns embedded within time series. This capability motivates a detailed examination of its various aspects in the context of standard dynamical systems. A notable example is the Bak-Tang-Wiesenfeld (BTW) sandpile model, where \textit{rare events} (large avalanches) correspond to hubs in the graph and play a pivotal role in shaping its structure~\cite{yang2009visibility}. The BTW model is recognized as the first prototypical system exhibiting self-organized criticality (SOC), characterized by critical behavior emerging without the need for external parameter tuning~\cite{bak1987self}. This model simulates avalanches triggered by energy redistribution in a critical state. Features such as the separation of time scales, non-linear threshold dynamics, and bursty behavior offer promising connections to real SOC systems, including earthquakes~\cite{sornette1989self, olami1992self}, rainfall~\cite{bove2006complexity}, clouds~\cite{najafi2021self, cheraghalizadeh2024simulating}, and neuronal activities~\cite{levina2007dynamical, bonachela2010self}. Several properties of the BTW model are now well understood, such as its connection to the $q \rightarrow 0$ Potts model~\cite{saleur1987exact}, spanning trees, $c=0$ conformal field theories~\cite{MAJUMDAR1992129}, various height probabilities~\cite{majumdar1991height}, and the $W$-algebra structure in its ghost field theory~\cite{mahieu2001scaling}. However, identifying the universality classes of sandpiles, based on their critical exponents, remains a significant challenge~\cite{pruessner2012self}. A crucial question in this context is to understand how highly active avalanches influence the system's dynamics. This issue has been effectively addressed using the visibility graph technique, which revealed connections between the BTW model and scale-free complex networks~\cite{yang2009visibility}. \\

Despite the growing interest in visibility graphs, most studies focus on lower-order characteristics, such as local node connectivity and degree centrality, which primarily capture short-range temporal correlations. However, SOC systems like the BTW model also exhibit long-range interactions and multiscale behavior, which are best understood through higher-order topological features. These features can be studied using advanced tools from algebraic topology, such as \textit{simplicial complexes} and \textit{persistent homology}, which reveal the presence of loops, voids, and other higher-dimensional structures within the visibility graph. These higher-order features provide insights into how avalanches of different sizes interact over time and contribute to the system’s self-organization. Topological perspective of complex networks has recently been highlighted, which, instead of local (node-dependent) properties of complex networks, focuses on the global properties~\cite{wasserman2018topological}. In this paper, we propose a comprehensive study of both the lower-order and higher-order connectivity of visibility graphs constructed from the time series of sandpile avalanches. By analyzing standard graph metrics like degree and betweenness centralities, we explore local connectivity properties that reflect small-scale interactions within the system. In parallel, we utilize topological methods to extract higher-order patterns that represent global interactions and long-range dependencies in the system evolution. The novelty of this work lies in its integration of graph theory and topological data analysis (TDA) to provide a multi-scale perspective on the dynamics of SOC systems. We aim to answer fundamental questions regarding how local toppling events influence global dynamics and whether visibility graphs can be used as a diagnostic tool to identify self-organized critical behavior in complex systems.

The rest of the paper is structured as follows. In Section~\ref{SEC:tda}, we present the mathematical formulation of higher-order graph connectivity. Sections~\ref{Sec:VG} and~\ref{Sec:BTW} provide a detailed description of the construction of the visibility graphs from time series and the BTW model, respectively. In Section~\ref{Sec:results}, we analyze the lower-order connectivity using standard graph metrics, followed by an extensive calculation of the higher-order topological features using simplicial complexes and persistent homology. Finally, in Section~\ref{Sec:conclusion}, we summarize the key findings and outline future directions for research in the topological analysis of complex networks. 

\section{Topological Analysis of Complex Networks}\label{SEC:tda}
Statistical features such as degree centrality, betweenness, and average shortest path length have been extensively employed to classify complex networks into universal categories like scale-free~\cite{barabasi2003scale,Adami2024centrality}, small-world~\cite{watts1998collective}, or random~\cite{erdos1959random} by comparing these metrics across different network topologies. While these features have been instrumental in characterizing networks, they primarily capture global and local structural properties, which may overlook deeper, more fundamental aspects of the network's underlying topology. Topological analysis offers an alternative, more robust framework for understanding the structure of complex networks by focusing on properties that remain invariant under continuous deformations.

Topology, which is connected to the properties preserved under deformation (such as stretching or bending but not tearing), provides a powerful lens for comparing and classifying networks beyond traditional metrics. In this context, topological invariants—such as the number of connected components (in the zeroth dimension), the number of loops or cycles (in the first dimension), and higher-dimensional features like voids (in the second dimension)—allow for a more fundamental classification of network structures. These topological invariants are critical in identifying the homological properties of a network, which provide insights into its structural robustness and connectivity patterns. Although complex networks are not inherently geometric objects, they can be treated as \textit{topological spaces}, allowing us to apply tools from algebraic topology, such as homology, to capture their multi-scale features. The challenge, however, lies in directly computing the homology groups of these networks due to their irregular and often high-dimensional nature. To address this, one can approximate the network with a \textit{simplicial complex} which is homeomorphic to the original network, meaning it retains its essential topological features. Homology provides a classification of the network's features across different dimensions: for example, zeroth homology identifies connected components, first homology captures loops, and higher homology reveals higher-dimensional voids. This topological perspective not only complements traditional statistical measures but can also uncover hidden structures and relationships within complex networks that are otherwise difficult to detect using conventional graph-theoretic methods. Below is an overview of the main mathematical concepts involved in the TDA of complex networks.

\subsection{Simplicial Complexes in Networks
}
To analyze the topology of complex networks, one approximate it with a simplicial complex,  a combinatorial representation that preserves the topological essence of the network built from simplexes (points, edges, triangles, and higher-dimensional analogs). In this context: 0-simplexes are the nodes (vertices) of the network. 1-simplexes are the edges between nodes.
Higher-dimensional simplexes (such as triangles, or 2-simplexes, and tetrahedra, or 3-simplexes) are related to higher order connections involving more than two nodes. The following definitions are crucial in our discussions: 
\begin{itemize}
    \item The $n$-simplex denoted by $\sigma_n=(v_0\: v_1\: ...\: v_n)$ in a space with dimension $\mathbb{R}^n$ is formally defined as
\begin{equation}
	\sigma_n = \{x\in \mathbb{R}^n \, | \, x=\sum^{n}_{i=0}\lambda_iv_i, \, \lambda_i\geq0, \, \sum^{n}_{i=0}\lambda_i =1\},
\end{equation}
where $\left\lbrace v_i\right\rbrace_{i=1}^n $ represents the set of nodes, and must be geometrically independent in order to represent an $n$-dimensional object. The $\lambda_i$s are some coefficients that satisfy the conditions of the above equation.

\item A \textit{$p$-face} of a $n$-simplex is a simplex itself made by $p+1$ points and it is a subset of the $n$-simplex such that $0\leq p\leq n$. 

\item A simplicial complex $K$ is a set whose elements are the simplexes with different dimensions. The dimension of a simplicial complex is defined to be the dimension of its largest simplex.

\end{itemize}
 
 Two conditions are necessary in order to make a simplicial complex which is going to be homeomorphism to the underlying topological object;
\begin{itemize}
	\item Every face of each simplex of a simplicial complex $K$ also belongs to $K$.
	\item The intersection of any pair of two simplexes in $K$ is either empty or a face of those two simplexes.
\end{itemize}

\subsection{Homology Groups and Betti Numbers}
Homology provides a way to quantify the topological features of a network at different dimensions. A $d$-dimensional homology focuses on $d$-dimensional voids; when $d=0$ and $d=1$ concern the number of connected components and the cycles or loops in the network respectively, while higher-dimensional homology describes voids in higher-order simplicial complexes. In this regard, the \textit{Betti numbers} ($\beta_d$) quantify the number of independent topological features: The 0th Betti number ($\beta_0$) counts the number of connected components, while $\beta_1$ counts the number of independent loops.\\

In order to obtain the homology groups of a simplicial complex that is homeomorphic to a graph, one has to understand what \textit{chain groups}, \textit{cycle groups} and \textit{boundry groups} of a simplicial complex are. Let $K$ be our $n$-dimensional simplicial complex. The $d$-chain group $C_d(K)$, $0\leq d\leq n$, with the addition operation is the set of all $d$-chains denoted by $c$ and is represented as 
\begin{equation}
	C_d(K) = \lbrace c \: \vert \: c=\sum^{N_d}_{i=1}c_i\sigma^i_{d}, \: c_i\in \mathbb{Z} \rbrace
\end{equation}
where $N_d$ is the total number of $d$-simplexes available in the simplicial complex $K$, and $\sigma_d^i$ is the $i$th $d$-dimensional simplex in $K$.
 The unit element of the $d$-chain group $C_d(K)$ is $0 = \sum_i 0.\sigma^i_{d}$. Also, each element $c$ has an inverse element of $-c = \sum^{N_d}_{i=1}(-c_i)\sigma^i_{d}$. The $C_d(K)$ has the properties of free Abelian groups, and its rank is $N_d$.

The $d$-cycle group denoted by $Z_d(K)$ is defined as 
\begin{equation}
	Z_d(K)=\lbrace c \: \vert \: \partial_dc=0, \: c\in C_d(K)\rbrace
	\label{eqt_r_cycle}
\end{equation}
where $\partial_d$ is called the boundary operater and it acts on simplexes as
\begin{equation}
	\partial_d\sigma_d=\sum^d_{i=0}(-1)^i(v_0\: v_1\: ... \: v_i\: ...\: v_n)
	\label{eqt_bndr_operator}
\end{equation}
where the point $v_i$ is ignored. The $d$-cycle group is a subgroup of $d$-chain group. The elements of the $d$-cycle group are called $d$-cycles. By combining the Eqs. (\ref{eqt_r_cycle}) and (\ref{eqt_bndr_operator}) one can find that the boundary operator $\partial_d$ is actually a map from $C_d(K)$ to $C_{d-1}(K)$
\begin{equation}
	\partial_d : C_d(K) \to C_{d-1}(K).
\end{equation}
The $d$-boundary group is defined as 
\begin{equation}
	B_d(K)= \lbrace c\: \vert \: c=\partial_{d+1}c', \: c'\in C_{d+1}(K)\rbrace.
\end{equation}
For an $n$-dimensional simplicial complex $K$ the $d$th homology group $H_d(K)$, $0 \leq d \leq n$, associated with $K$ is defined by
\begin{equation}
	H_d(K)\equiv Z_d(K)/B_d(K).
\end{equation}
which is topological invariant. For a simplicial complex $K$, the $d$th Betti number, $\beta_d(K)$, is defined by
\begin{equation}
	\beta_d(K)\equiv \mathrm{rank}\, H_d(K).
\end{equation}

\subsection{Filtration and Persistent Homology}~\label{SEC:persistent_homology}

\begin{figure}
		\includegraphics[width=0.48\textwidth]{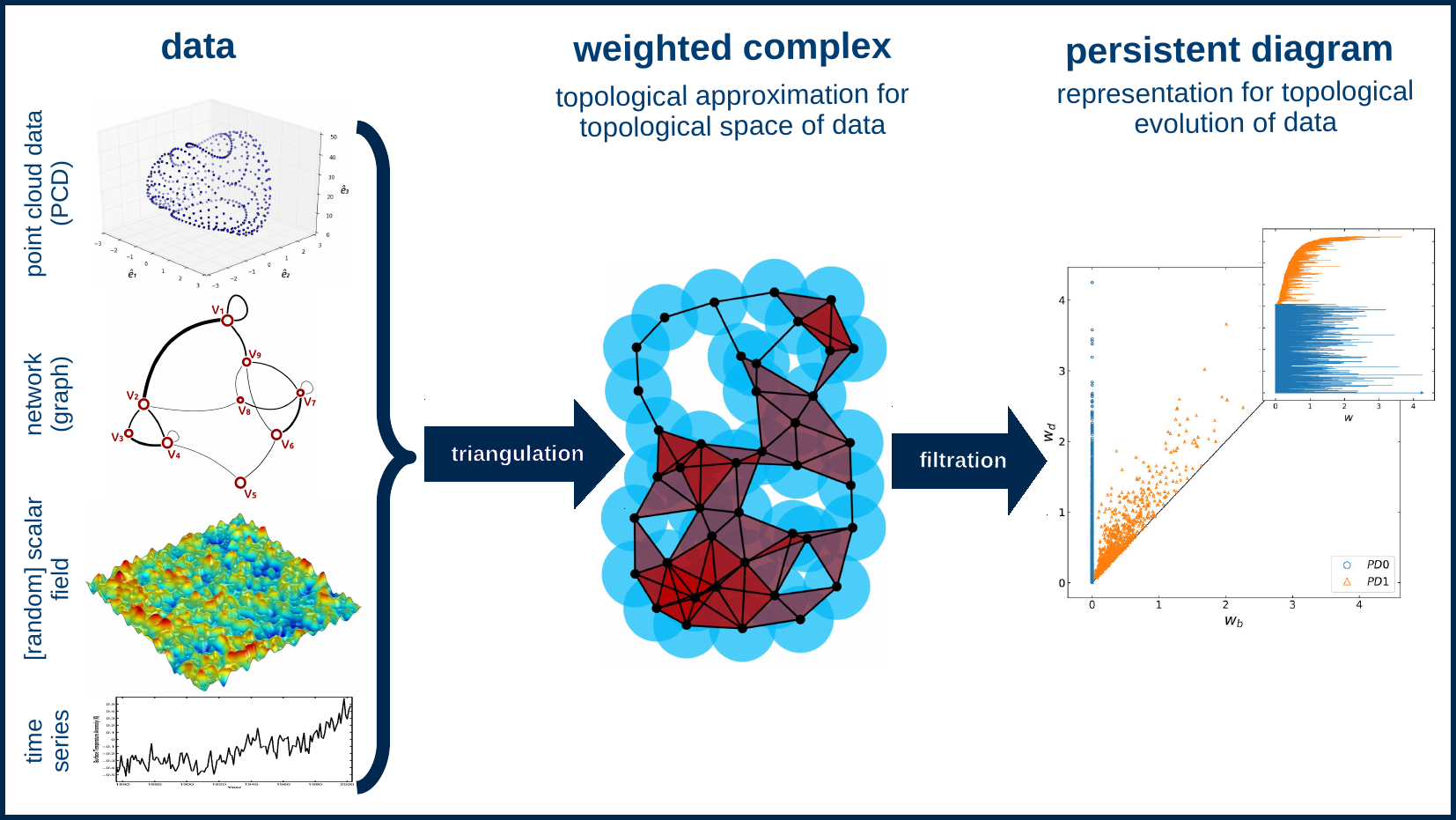}
	\caption{TDA technique: It transforms various kinds of data that depend on some parameter into a weighted simplicial complex, which is a topological approximation of the phase space of the systems. This technique finds homology generators (topological invariants) of the complexes in the filtration process and summarizes them as persistent pairs in a persistent diagram to capture how the topological information in the dataset changes.}
	\label{fig:TDA_pipeline}
\end{figure}
\textit{Filtrations} are used to analyze how the topology of a network evolves as the network is built up step by step~\cite{edelsbrunner2002topological}. In TDA, a filtration is a sequence of nested simplicial complexes where additional edges or simplexes are added based on some criteria, such as a distance threshold or network weight. This allows for tracking the birth and death of topological features at different scales. Let $w$ represent the filtration parameter. At each level, one can construct the corresponding simplicial complex $K(w)$ and find the topological features in terms of that special value. Then $w$ is increased to some $w+\Delta w$ at the next step. This increasing of the parameter will add some new simplexes to $K(w)$ and as a result, it will be updated to a new simplicial complex $K(w+\Delta w)$. All the elements (simplixes) of $K(w)$ are also the elements of $K(w+\Delta w)$.  Accordingly, 
\begin{equation}
    K(w) \subset K(w+\Delta w)
    \label{eq:filtration1}
\end{equation}

Persistent homology is a method in TDA that can be used to compute topological invariants of the network and calculate how long they can exist (persist) when the filtration parameter varies. As the filtration progresses, topological features such as connected components and loops can appear (birth) and disappear (death). Features that persist longer are more likely to be real features of the space, not just caused by how we sample, measure, or choose parameters \cite{carlsson2009topology}.

To discover the hidden topological structure of a network, persistent homology observes the changes of homology generators of dimension $d$ ($d$-dimensional topological holes), denoted by $h_{d}$, by adjusting the filtration parameter and assigning an ordered pair $(w_{\rm{birth}}^{(h_d)},w_{\rm{death}}^{(h_d)})$, known as persistence pair, indicating the filtration parameter when the hole $h_{d}$ is born and dies. These ordered pairs may be plotted as points on a two dimensional Euclidean space called \textit{persistence diagram} \cite{edelsbrunner2002topological} or they can be represented on a one dimensional diagram  called \textit{persistence barcode} \cite{carlsson2004persistence} where the threshold parameter plays the role of the persistence barcode's variable. The start and end points of the \textit{bar}s of each $h_d$ are determined by $w_{\rm{birth}}^{(h_d)}$ and $w_{\rm{death}}^{(h_d)}$, respectively. Figure \ref{fig:TDA_pipeline} summarizes the TDA technique for different types of data.

Another quantity that one may employ in the TDA is the lifetime of a topological hole $h_{d}$, which is the difference between the threshold when the hole is born and when it dies, $\ell^{(h_d)} \equiv w_{\rm{death}}^{(h_d)} - w_{\rm{birth}}^{(h_d)}$. Also, persistence entropy of $d$th persistence diagram is the Shannon entropy of lifetime of persistence pairs in $d$th persistence diagram. By introducing $\mathcal{L} \equiv \displaystyle \sum_{w^{(h_{d})} \in \mathcal{M}(PD_d)} \ell^{(h_d)}$ we have
\begin{equation}
	PE_d = - \displaystyle \sum_{w^{(h_{d})} \in \mathcal{M}(PD_d)} \frac{\ell^{(h_d)}}{\mathcal{L}} \quad \log \Bigr[ \frac{\ell^{(h_d)}}{\mathcal{L}} \Bigr]
 \label{eq:lifetime}
\end{equation}
where $PD_d(K(w))\equiv (\mathcal{M,N})$ with $\mathcal{M}\equiv \{ w^{(h_{d})} | h_d\in H_d(K(w))\}$ and $\mathcal{N}:\mathcal{M}\to \mathbb{N}$ is the count function.

\section{Visibility Graph}\label{Sec:VG}
Let us consider a time series $S(N) \equiv \left\lbrace s(t_i)\right\rbrace_{i=1}^N$, where $s(t_i)$ represents the time series variable, a real number for each $i$, and $\left\lbrace t_i\right\rbrace_{i=1}^N$ denotes the set of measurement times. To associate it with activity, we focus on the case where $s(t_i) \geq 0$, though the extension to more general cases is straightforward. This could represent activity recorded as a function of time in a dynamical system. By defining $s_i \equiv s(t_i)$, the visibility graph transformation involves mapping each data point $(t_i, s(t_i))$ to a graph node $(i, s_i)$. The resulting graph is $\mathcal{G} \equiv \left\lbrace \mathcal{V}, \mathcal{E} \right\rbrace$, where $\mathcal{V}$ is the set of nodes (times), and $\mathcal{E} \equiv \left\lbrace e_{ij} \right\rbrace_{i,j=1}^N$ is the set of edges determined by a geometric visibility criterion, which will be defined subsequently. The two main types of visibility graphs used in time series analysis are the \textit{natural} visibility graph and the \textit{horizontal} visibility graph. 
\begin{figure}[t]
  \centering
  \includegraphics[scale=0.95]{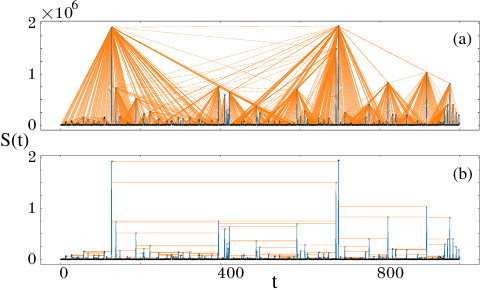}
      \caption{Visibility Graph created from the time series data by connecting every point (every bar) with all the other points that are visible from its top in natural (a) and horizontal (b) manner. In such a graphs, each node represents a data point in the same order as in the series, and two nodes are linked if their corresponding data points can see each other, meaning that there is no data point in between that is higher than the straight line joining them. The time-series used here has been captured from avalanche sizes happened on a square lattice of length $L=512$ of the dynamical BTW sandpile model. Here, the process of injecting grains has been repeated for as many as $10^3$ times.}
	\label{fig:NVG_HVG_time_series}
\end{figure}

In the natural visibility graph (which is considered in this paper), nodes are connected based on the geometric visibility between the corresponding points in the time series. Two points $(i ,s_i)$ and $(j ,s_j)$ are connected by an edge ($e_{ij}=1$) if and only if they are visible to each other, i.e. the line segment joining these points does not intersect any other intermediate data points. As an instance see  Fig.~\ref{fig:NVG_HVG_time_series}(a) where the visible connections are identified using a direct line. Mathematically, two points $i$ and $j$ have visibility (connection) if, for every $k$ between $i$ and $j$, the following condition holds:
\begin{equation}
    s_k<s_i+\frac{s_j-s_i}{j-i}(k-i),\ \forall k\in [i,j].
    \label{Eq:edges}
\end{equation}

This condition ensures that no data point $s_k$ obstructs the visibility between $s_i$ and $s_j$. The resulting natural visibility graph captures the overall temporal structure of time series, including correlations between small and large activities across time.

\begin{figure*}[t]
  \centering
  \includegraphics{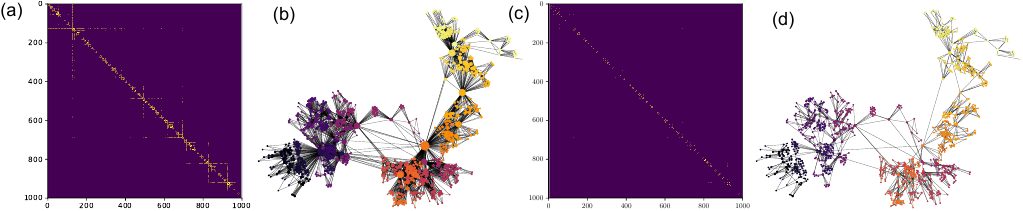}
  \caption{(a) and (b) are adjacency matrix and a visualization of the corresponding graph of a natural visibility graph, while (c) and (d) are those of a horizontal visibility graph represented in Fig.\ref{fig:NVG_HVG_time_series}. The size and color in (b) and (d) reveal the degree and the birth-time of the nodes. The bigger the node, the higher its degree value. The birth-time indicates when the node was added to the graph. The darker the node, the older it is.}
	\label{fig:AMs_VGs}
\end{figure*}

The horizontal visibility graph is a simpler variant of the natural visibility graph, where two points $(i ,s_i)$ and $(j ,s_j)$ are connected if and only if no intermediate point has a greater value than both $s_i$ and $s_j$; Fig.~\ref{fig:NVG_HVG_time_series}(b). The horizontal visibility graph provides a more localized view of the time series dynamics, focusing on short-term temporal correlations, while maintaining many of the key structural properties of the original time series. Figures~\ref{fig:AMs_VGs}(a) and~\ref{fig:AMs_VGs}(c) present the adjacency matrices of two samples of natural and horizontal visibility graphs, respectively. These matrices are visually represented as graphs in Figs.~\ref{fig:AMs_VGs}(b) and~\ref{fig:AMs_VGs}(d).

In this study, we assign weights to the edges in our natural visibility graph to further analyze the higher-order topological features of the system. These weights will serve as the filtration parameter for calculating topological invariants using the Vietoris-Rips filtration method (see Section~\ref{SEC:persistent_homology}). Based on the visibility condition defined by Eq.~\ref{Eq:edges}, we assign the weights to the edges in a straightforward manner, thereby constructing a weighted natural visibility graph:
\begin{widetext}
	\begin{equation}
	w_{ij} = \left\{
	\begin{array}{l l}
	\Bigr{|}Q_{ij} \Bigr{|} \qquad ; \qquad \Bigr{|}j-i \Bigr{|} = 1 \\
	\Bigr\{ \displaystyle \prod_{k=i+1}^{j-1} \Theta \Bigr{[} Q_{ij} - Q_{ik} \Bigr{]}
	\Bigr( Q_{ij} - Q_{ik} \Bigr) \Bigr\}^{1/(j-i-1)}  ; \qquad \Bigr{|}j-i \Bigr{|} > 1,
	\end{array} \right.
	\label{Eq:weights}
	\end{equation}
\end{widetext}
where $Q_{ij}= \frac{s_j - s_i}{j - i}$ and $\Theta$ is the step function, i.e. $\Theta(x)=1$ if $x\geq 0$ and zero otherwise. We use this type of weight, indicating the \textit{quality of the visibility}, for our analysis. This weight has been used before for studying fractional Gaussian noise, which reveals some hidden exponents related to the Hurst exponent~\cite{masoomy2023relation}. This weighted approach allows for a more detailed examination of the topological structure and connectivity properties of the visibility graph at different scales.\\

The visibility graph approach not only preserves the inherent complexity of the original time series but also provides a new perspective on SOC systems by revealing topological properties such as degree distribution, clustering, and network connectivity. These graph-based metrics can uncover key aspects of the system's dynamics, from the frequency and size of avalanches (lower-order connectivity) to the emergence of large-scale structures and hierarchical patterns (higher-order connectivity).

\section{Bak-Tang-Weisenfeld Model}\label{Sec:BTW}

The BTW model is a paradigmatic model for self-organized criticality (SOC) that provides a mathematical framework for systems that naturally evolve into a critical state. In this state, small perturbations may lead to large-scale, system-wide events, i.e. rare avalanches. We consider the BTW model on a two-dimensional lattice. The dynamics of the system is based on the redistribution of the grains or energy units throughout the system when the local configuration exceeds a critical threshold.

\begin{figure}
    \centering
    \includegraphics{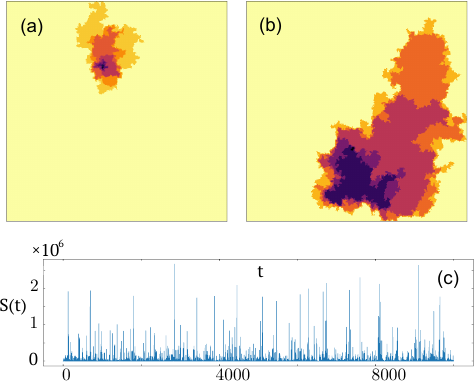}
    \caption{Two avalanches with different sizes are shown in (a) and (b) happening on a square lattice of size $L=512$ according to the BTW model. (c) Time-series of avalanche sizes taken placed on a square lattice of size $L=512$ simulated by using the dynamics of the BTW sandpile model. Here, the process of injecting grains has been repeated for as many as $10^4$ times.}
    \label{fig:btw_time_series}
\end{figure}

Consider a $L\times L$ square lattice and attribute a local height of the number of sand grains or energy units to each point, denoted by an integer number $z(m,n)$. The model is also endowed by a threshold $z_{\text{crit}}$ (considered to be 4 in this paper): a site with a height higher than $z_{\text{crit}}$ is called \textit{unstable} and should \textit{topple}: it gives one grain to each of its neighbors. The system evolves through the addition of grains at randomly chosen sites. At any given time step, the height of each point is updated as grains are added to the system. If the random site to add the new sand grain is ($m_0,n_0$), i.e. $h(m_0,n_0)\to h(m_0,n_0)+1$ then the toppling is mathematically described as:
\begin{equation*}
   \text{If}\ z(m_0,n_0) > z_{\text{crit}} \Rightarrow z(m_0,n_0) \to z(m_0,n_0)-4     
\end{equation*}
where four grains are redistributed to the nearest neighbors:
\begin{equation*}
    z(m_0\pm 1,n_0\pm 1) \to z(m_0\pm 1,n_0\pm 1) + 1.
\end{equation*}
Note that for the boundary site, one or two sand grains may dissipate. This redistribution can trigger toppling events at neighboring sites, leading to a domino of topplings (with $(m_0,n_0)$ replaced with new unstable points in the above equation). This cascade of topplings is refer to as an \textit{avalanche}. Each sand grain injection is considered as a time step $t_i$: with any injection the time step proceeds by one unit, which may lead to and avalanche with the size $s(t_i)$, which is considered to be the time series under investigation. The size of an avalanche is defined as the total number of toppling events triggered by the addition of a single grain. Two small and large avalanche samples of the BTW are shown in Figs.~\ref{fig:btw_time_series}(a) and~\ref{fig:btw_time_series}(b), respectively. The system stabilizes when no sites exceed the critical threshold $z_{\text{crit}}$ and we can continue the dynamic by adding the next grain. At the \textit{boundaries} of the lattice, if grains are redistributed outside the lattice, they are lost from the system. This can be understood as imposing \textit{open boundary conditions}, which ensure that the system remains finite and avoids unbounded growth of the pile heights. These boundary conditions play a key role in maintaining the critical state of the system by allowing excess grains to dissipate.

Mathematically, SOC is reflected in the system's long-range correlations and the absence of a characteristic avalanche size. The power-law behavior indicates that the system exhibits scale invariance, with avalanches of all sizes possible. This can be described by scaling relations~\cite{najafi2012avalanche,najafi2016bak,najafi2014bak1,Dhar1990Self,dashti2015statistical} and renormalization group~\cite{ivashkevich1999dynamical,lin2002renormalization} techniques, connecting the BTW model to broader theoretical frameworks in statistical physics~\cite{najafi2021some}.

The distribution of avalanche sizes follows a power-law distribution, which is characteristic of self-organized critical systems:
\begin{equation}
    P(s)\sim s^{-\tau}
\end{equation}
where 
$P(s)$ denotes the distribution function of an avalanche of size $s$, and $\tau$ is a critical exponent that depends on the dimensionality and geometry of the lattice. For the 2D BTW model, $\tau$ has been numerically estimated to be approximately 1.25~\cite{manna1991two}.

In this paper, we consider the process of constructing visibility graphs from the avalanche time series generated by the BTW model to investigate the complex temporal dynamics of avalanches using graph-theoretical tools. The visibility graph for sandpiles was examined first in~\cite{masoomy2023relation}. In this case the time series is associated with the size of each local activity of the $i$th avalanche, i.e. $s_i$ occurred at time step $t_i\equiv i$. We record only time steps that result in toppling, and $N$ is the total number of avalanches over the observation period. Figure~\ref{fig:btw_time_series}(c) is a sample of avalanche time series with a crucial (most visible) role or \textit{rare events} (typical for self-organized critical systems) corresponding to the hubs in the visibility graphs.

\section{Results}\label{Sec:results}
We simulated the BTW sandpile model on square lattices of varying sizes ($L=$ 64, 128, 256, 512, 1024, 2048). For each lattice, we generated visibility graphs from time series of different lengths, $N$s, representing the number of observed avalanches. To ensure statistical reliability, each configuration was simulated $10^4$ times. For lower-order network features like degree and betweenness centrality, we used the NetworkX Python library~\cite{hagberg2008exploring}. To compute persistent homology and extract higher-order topological features, we employed the Dionysus Python library~\cite{dmitriydionysus}. 

In the following, we present our findings on lower-order features, including degree distributions and betweenness centralities. We then discuss the results of our higher-order topological analysis.

\subsection{Lower-Order Connectivity Properties}
\begin{figure}[t]
  \centering
  \includegraphics{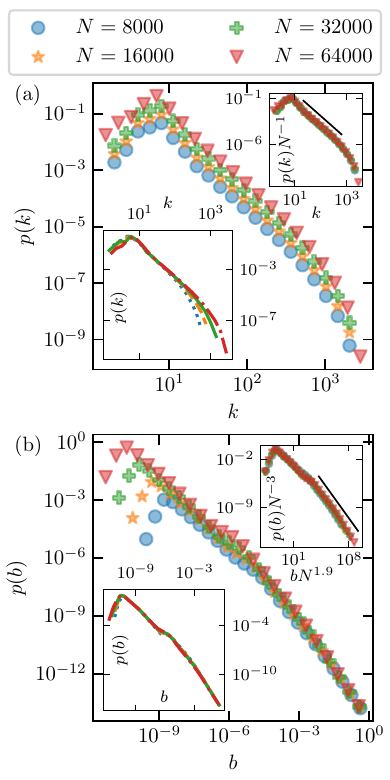}

      \caption{Probability distribution function of (a) degree and (b) betweenness centrality of the binary (unweighted) graphs of the BTW model for different network sizes. Both of these centralities reveal power law behavior. The upper right insets in both cases are the outcome of data collapse of the corresponding PDFs in the self-similar coordinates of $k$ and $p(k)N^{-1}$ in PDF of degree centrality, and $bN^{1.9}$ and $p(b)N^{-3}$ in betweenness centrality. We obtain the slopes of $2.50\pm 0.02$ and $1.585\pm 0.008$ for degree and betweenness centralities, respectively. In the lower left insets of these plots, we examine the corresponding PDFs as we change the lattice length ($L=256$ dotted line, $L=512$ dashed line, $L=1024$ solid line and $L=2048$ dashed dotted line) setting $N$ fixed as 64000.}
	\label{fig:localstat}
\end{figure}
Here, we analyze the lower-order connectivity properties of the visibility graphs derived from the time series of the BTW sandpile model. Specifically, we focus on two central measures of network connectivity for the binary visibility graph: degree centrality and betweenness centrality. These measures provide insights into the local and global structural characteristics of the visibility graph, allowing us to explore the interplay between the dynamics of sandpile avalanches and the topological structure of their associated visibility graphs.

\textit{Degree centrality} is one of the most fundamental measures of connectivity in a network, defined as the number of edges connected to a vertex, denoted by $k_i=\sum_{j}a_{ij}$, where $A=\left\lbrace a_{ij}\right\rbrace_{i,j=1}^N $ is the adjacency matrix, i.e. $a_{ij}=1$ if $i$ and $j$ are connected, and zero otherwise. In the context of a visibility graph, the degree of a node reflects the number of other avalanche events (represented as nodes) that are directly visible from that node, based on the visibility condition described in Eq.~\ref{Eq:edges}. This feature is a local measure of connectivity, capturing the immediate neighborhood of an avalanche event.\\

For the BTW sandpile model, the degree distribution of the visibility graphs exhibits characteristics typical of scale-free networks; Fig~\ref{fig:localstat}(a). The tail of the distribution follows a power law, $P(k)\sim k^{-\gamma}$, where $k$ is the degree of a node and $\gamma$ is the scaling exponent. This indicates that most avalanche events have a low degree, while a few are highly connected associated with the rare events, acting as hubs in the network. As a result, the visibility graph captures the long-range temporal correlations inherent in the sandpile's self-organized criticality.

\textit{Betweenness centrality} measures the extent to which a node lies on the shortest paths between other nodes in the network. This centrality is crucial for understanding the global structure of the visibility graph, as it highlights nodes that serve as bridges or bottlenecks in the network’s connectivity. To define the betweenness centrality, let us denote the shortest path length between nodes which is set to $N$ when they are not connected by any path. Then, the betweenness centrality of a node $i$  is given by
\begin{equation}
b_i={\sum_{j\neq i}}{\sum_{k\neq i}}\frac{n_{jk}(i)}{n_{jk}},
\end{equation}
where $n_{jk}$ is the number of shortest paths between $j$ and $k$, whereas $n_{jk}(i)$ is the number of shortest paths between $j$ and $k$ that go through node $i$. In the visibility graph of the BTW model, nodes with high betweenness centrality represent avalanche events that play a key role in connecting different clusters of events, thereby mediating the flow of information (or influence) through the network.

For the BTW visibility graph, betweenness centrality, as one can see in the Fig.~\ref{fig:localstat}(b), tends to follow a broad distribution, reflecting the heterogeneous nature of avalanche dynamics. A small number of nodes exhibit very high betweenness centrality, corresponding to rare, large-scale avalanches that connect otherwise disparate regions of the time series. These avalanche events are crucial for maintaining the overall connectivity of the graph, as their removal would significantly alter the structure of the network, possibly fragmenting it into smaller disconnected components. The high betweenness centrality of certain nodes also reveals important insights into the time correlations within the BTW model. Large avalanches, while infrequent, act as central connectors across time, linking clusters of smaller avalanches that would otherwise remain isolated in the visibility graph. \\

There are two finite size effects in this problem: one is the BTW system size $L$, and the maximum time $N$ considered in the simulations, the latter identifying the size of network. As is seen in Figs.~\ref{fig:localstat}(a) and~\ref{fig:localstat}(b) the probability distribution functions (PDFs) of degree and betweenness (considered for different $N$ and $L$ values) follow power-law relationship for at least two decades. Both of the functions satisfy the following finite size scaling relation in terms of $N$ ($x\equiv k,b$):
\begin{equation}
p_x(x)= N^{\alpha_x}F_x\left(xN^{\tau_x} \right)=x^{-\gamma_x}G_x\left(xN^{\tau_x}\right)
\label{Eq:FSS}
\end{equation} 
where $\gamma_x\equiv\frac{\alpha_x}{\tau_x}$, and $F_x(y)$ and $F_x(y)$ are some universal functions related via the relation $F_x(y)=y^{-\gamma_x}G_x(y)$, with the restriction $\lim_{y\gg 1}G(y)=const.$, guaranteeing the power-law behavior. Based on our data collapse analysis presented in the upper insets of these figure, and within the error bars we obtained, we propose the following relation
\begin{equation}
\alpha_x=\tau_x+1, \ x=k,b, 
\end{equation}
and also $\tau_k\approx 0$, and $\tau_b\approx 2$ (the exact values are reported in the figure). While $\gamma_b$ can readily be determined using the relation $\gamma_b=1+\frac{1}{\tau_b}$ ($\gamma_{b}(N=64000)=1.585\pm 0.008$), the determination of $\gamma_k$ has challenges since $\tau_k$ is almost zero. By fitting the $p(k)$ in the range $k\in [10-200]$ we obtain $\gamma_k=2.50\pm0.02$, while we are not sure if this power-law dependence lasts for larger $k$ scales. The finite size effects regarding the BTW system size $L$ for $N=64000$ are presented in the lower insets in both graphs. While for $k$ the graphs show a dependence on $L$, no such a dependence is observed for $b$.

\subsection{Higher-Order Connectivity Properties} \label{sec:topological_properties}
\begin{figure*}[t]
  \centering
  \includegraphics{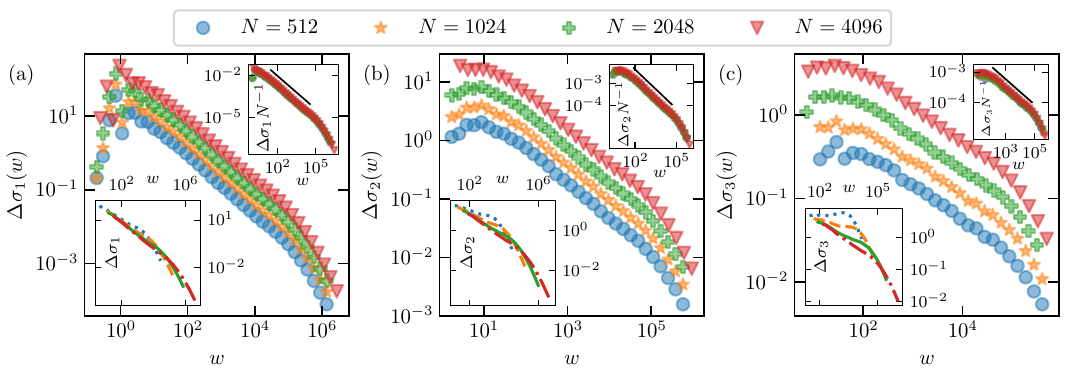}
      \caption{Weight-dependent number of (a) links (1-dimensional simplexes), (b) triangles (2-dimensional simplexes), and tetrahedrons (3-dimensional simplexes) in weighted natural visibility graph of different network sizes. The upper right inset in each of these figures is the outcome of data collapse of their main figures in the self-similar coordinates by multiplying the vertical axis by $N^{-1}$. The values of slopes are obtained as $\gamma_{\sigma_1}=0.795\pm0.006$, $\gamma_{\sigma_2}=0.602\pm0.009$ and $\gamma_{\sigma_3}=0.422\pm0.008$. In the lower-left insets of these plots, we examine how the weighted natural visibility graph of a time-series affects the number of simplexes as we change the lattice length ($L=256$ dotted line, $L=512$ dashed line, $L=1024$ solid line and $L=2048$ dashed dotted line) and the filtration parameter keeping $N$ fixed as 4096.}
	\label{fig:simplexes}
\end{figure*}
In Section~\ref{SEC:persistent_homology}, we emphasized the role of filtration in computing persistent homology for topological spaces. Several filtration techniques have been introduced~\cite{horak2009persistent, chowdhury2016persistent, rieck2017clique} to compute persistent homology in various types of networks, such as weighted, directed, and dynamic networks. In this work, we utilize the Vietoris-Rips filtration~\cite{zomorodian2005}, a common approach for analyzing the persistence of homological features in undirected weighted networks. Here, the filtration parameter is taken as the edge weights, denoted by $w$ throughout the paper.

The Vietoris-Rips filtration process begins by arranging the edge weights, determined by Eq.~\ref{Eq:weights}, from a minimum $w_{\text{min}} \equiv 0$ to a maximum value $w_{\text{max}}$, increasing by increments of $\Delta w$. At each step, a corresponding simplicial complex is formed, referred to as the \textit{clique complex} denoted by $Cl(w)$. Initially, when $w_{\text{min}} = 0$, the clique complex consists solely of the disconnected vertex set. As the filtration progresses, edges with weights less than or equal to the current filtration value $w$ are added, causing the clique complex to expand. With each increment, the new clique complex contains the one from the previous step, which can be described as:
\begin{equation}
    Cl(G_w) \hookrightarrow Cl(G_{w + \Delta w})
    \label{eq:filtration2}
\end{equation}
similar to Eq.~(\ref{eq:filtration1}), where $G$ is the underlying graph.\\

We begin by analyzing the statistics of higher-order structures—specifically, the $1$-simplexes (links), $2$-simplexes (triangles), and $3$-simplexes (tetrahedrons)—across various network sizes, $N$, in terms of the parameter $w$, as illustrated in Figs.~\ref{fig:simplexes}(a),~\ref{fig:simplexes}(b) and~\ref{fig:simplexes}(c). It is important to note that, as shown by Eqs.~\ref{eq:filtration1} and~\ref{eq:filtration2}, the count of $d$-dimensional simplexes increases as a function of the filtration parameter $w$. Therefore, to determine the precise number of simplexes generated by changing $w$ to $w+\Delta w$, one must calculate the difference in their counts over this interval. Thus, we define
\begin{equation}
    \Delta\sigma_d = \sigma_d(w+\Delta w) - \sigma_d(w).
\end{equation}
where $d$ refers to the dimension of the simplex. Our analysis on this function, integrated in Fig.~\ref{fig:simplexes} (see the upper insets) reveals that
\begin{equation}
\Delta\sigma_d=N\mathcal{F}_d\left(w\right),
\label{Eq:sigma}
\end{equation}
i.e. $N^{-1}\Delta\sigma_d$ is independent of $N$. The fitting of $N^{-1}\Delta\sigma_d$ in the interval $w\in[10^1-10^4]$ reveals a power-law behavior as follows
\begin{equation}
    \mathcal{F}_d(w) \sim w^{-\gamma_{\sigma_d}},
    \label{Eq:sigma2}
\end{equation}
where $\gamma_{\sigma_d}$ is the corresponding exponent. We used the least square method to estimate the exponent. Our estimations are $\gamma_{\sigma_1} = 0.795 \pm 0.006$, $\gamma_{\sigma_2} = 0.602 \pm 0.009$ and $\gamma_{\sigma_3} = 0.422 \pm 0.008$. The lower insets show the $L$ dependence. We see that for the higher lattice sizes, the power-law fitting are better, Specifically for $d=3$, the power-law dependence is destroyed for lower $L$ values.

\begin{figure*}
\includegraphics{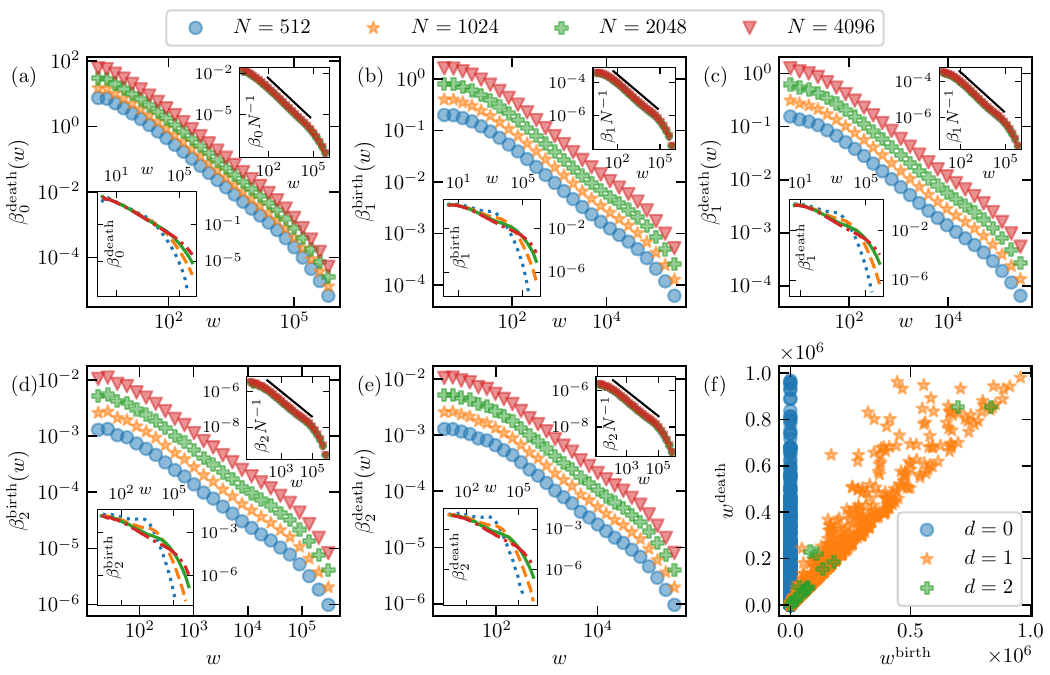}
\caption{\label{fig:betties} Betti numbers born in (b) first and (d) second dimensions and die in (a) zeroth, (c) first, and (e) second dimension as a function of the filtration parameter, $w$, for different sizes of the visibility graphs, $N$. These Betti numbers determine power law behavior in terms of $w$. The upper right inset in each of these plots shows that rescaling the vertical axis by $N^{-1}$ makes all the data in the main figure collapse into one curve. The values of the slopes are $\alpha_{0}^{\rm (death)} = 1.035 \pm 0.004$, $\alpha_{1}^{\rm (birth)} = 0.731\pm0.008$, $\alpha_{1}^{\rm (death)} = 0.742\pm0.008$, $\alpha_{2}^{\rm (birth)} = 0.715\pm0.007$, and $\alpha_{2}^{\rm (death)} = 0.70\pm0.01$. In the lower left insets in (a) till (e), we examine how the weighted natural visibility graph of a time-series affects the homology generators as we change the lattice length ($L=256$ dotted line, $L=512$ dashed line, $L=1024$ solid line and $L=2048$ dashed dotted line) and the filtration parameter. (f) Here, We show a persistent diagram of one sample of VG with the parameters of $L=2048$ and $N=4096$.}
\end{figure*}

When $w=w_{\text{min}}=0$, we will only have $\beta_0$ number which is equal with the total number of vertices while the higher dimensional Betti numbers are zero. As we increase the threshold parameter $w$,  other homology generators (topological loops and voids the number of which is indicated by $\beta_1$ and $\beta_2$, respectively) are born and the corresponding Betti numbers grow. At the same time, addition of simplexes (as a result of increasing the amount of $w$) can also cause the death of some other homology generators. Figure~\ref{fig:betties} shows various Betti numbers ($\beta_d$, $d=0,1$ and $2$) in terms of the the threshold $w$. In each $w$ the number of $d$-dimensional homological objects that are born (destroyed) is represented by $\beta_d^{\text{birth}}$ ($\beta_d^{\text{death}}$), which  follow the following scaling form for the visibility graph of sandpiles:
\begin{equation}
\beta_{d}^{(X)}(w) =N\mathcal{G}_d^{(X)}(w),
\end{equation}
where $X\equiv \text{birth}, \text{death}$. We see that for all $d$ values considered in this paper, and for both birth and death cases, this scaling is correct (see the upper insets). This demonstrates that the topological objects (reflected in the quantities $\Delta\sigma_d$ and $\beta_d$) per site (i.e. divided by $N$) is independent of system size. Additionally, in a wide $w$ range (at least four decades) the functions are governed by a power-law behavior: 
\begin{equation}
\mathcal{G}_d^{(X)}(w)\sim w^{-\alpha_{d}^{(X)}},
\end{equation} 
where $\alpha_{1}^{\text{birth}} = 0.731 \pm 0.008$ and $\alpha_{2}^{\text{birth}} = 0.715\pm 0.007$, and $\alpha_{0}^{\text{death}} = 1.035 \pm 0.004$, $\alpha_{1}^{\text{death}} = 0.742 \pm 0.008$, and $\alpha_{2}^{\text{death}} =0.70\pm0.01$.\\

\begin{figure}
\includegraphics{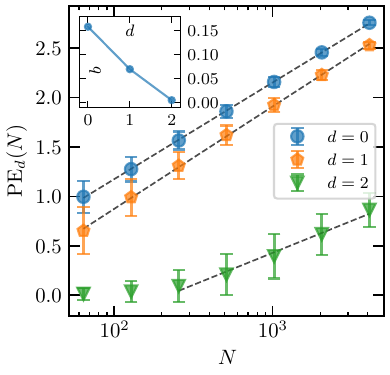}
\caption{\label{fig:PE} The persistent entropy of lifetime of $d$-dimensional holes increases logarithmically as the network size grows. The correspondent exponents for 0-holes, 1-holes and 2-holes are $0.159(4)$, $0.070(2)$ and $0.0046(6)$ respectively.}
\end{figure}

Figure~\ref{fig:betties}(f) shows a persistent homology diagram for $d=0,1$ and $2$ which is the set of pairs $(w^{\text{birth}},w^{\text{death}})$ for each single object. Since $\beta_0$ represents the number of connected components in the network, it is born once at $w=0$ and never dies for larger $w$ values. By analyzing the persistent homology of the visibility graphs, we extracted new exponents and scaling laws that characterize the temporal and topological structure of avalanches. To understand the physical implications of the 1-homology generator (loops), we consider a concave and a convex time series. One can easily see that for the concave time series, there is an all-to-all connection, while for the convex time series, only neighboring nodes are interconnected. In the latter case, there are only first neighbor connections, and no situation corresponds to a loop, since the first and the last nodes are not visible with each other. To have loops, one needs a set of nodes that are weakly visible to each other, so that the weight of most of them does not exceed the threshold in the filtration process. Two-dimensional and higher-dimensional features (voids, cavities) indicate even more complex patterns of interaction between avalanches, suggesting hierarchical or multi-scale structures in the visibility graph. These features reflect higher-order connectivity properties that go beyond simple pairwise connections between nodes. They capture global patterns of interaction that arise from the critical dynamics of the BTW model, offering a topological perspective on how avalanches interact over time.

The persistent entropy of the lifetime (introduced in the Eq.~\ref{eq:lifetime}) of the generators of 0-homology, 1-homology and 2-homology increases with a logarithmic trend, with the slopes of $0.159(4)$, $0.070(2)$ and $0.0046(6)$, respectively, as shown in the Fig. \ref{fig:PE}. We report all the  exponents obtained from calculating the exponents of the different statistical and topological features in this paper in the Table \ref{tab:exponents}.
\begin{table*}
\caption{\label{tab:exponents}Calculated exponents for different statistical and topological features in terms of the threshold parameter $w$ where the size of the lattice of the BTW model has been taken a constant value of $L=2048$.}
\begin{ruledtabular}
\begin{tabular}{ccccccccccc}
    exponent & $\gamma_{k}$ & $\gamma_{b}$ & $\gamma_{\sigma_1}$ & $\gamma_{\sigma_2}$ & $\gamma_{\sigma_3}$ & $\alpha_{0}^{\rm death}$ & $\alpha_{1}^{\rm birth}$ & $\alpha_{1}^{\rm death}$ &$\alpha_{2}^{\rm birth}$& $\alpha_{2}^{\rm death}$ \\ \hline
    value & $2.50(2)$ & $1.585(8)$ & $0.795(6)$ & $0.602(9)$ & $0.422(8)$ & $1.035(4)$ & $0.731(8)$ & $0.742(8)$ & $0.715(7)$ & $0.70(1)$ \\ 
\end{tabular}
\end{ruledtabular}	
\end{table*}

\section{Conclusion}\label{Sec:conclusion}
In this paper, we presented a comprehensive analysis of the lower-order and higher-order connectivity features of the visibility graph representation of sandpile models, with a focus on the BTW model of SOC. By transforming the time series of avalanches into visibility graphs, we explored the structural and topological characteristics of both local and global aspects of the system’s dynamics. Our results reveal several new scaling behaviors and critical exponents that provide deeper insights into the complex temporal correlations and multiscale organization inherent in SOC systems.

At the local (lower-order) level, we observed that standard graph-theoretic metrics such as degree and betweenness distributions exhibit power-law patterns that correspond to the scale-free behavior typical for SOC systems. The degree exponent of $\gamma_k=2.50\pm 0.02$ lies within the range $[2-3]$, typical for other scale-free networks. A novel scaling relation was proposed for the degree and the betweenness exponents, presented in Eq.~\ref{Eq:FSS}.

At the global (higher-order) level, we applied tools from TDA, such as simplicial complexes and persistent homology, to uncover higher-dimensional structures in the visibility graph. Our analysis revealed the presence of topological loops, voids, and other high-dimensional features that indicate long-range interactions and hierarchical patterns in the system's evolution. These higher-order features provide a new perspective on how large-scale avalanches propagate and influence the network, further enriching our understanding of the system’s global organization. Importantly, the higher-order connectivity analysis revealed new scaling behaviors that were not apparent from lower-order graph measures alone. Equation~\ref{Eq:sigma} shows the finite size scaling behavior of higher order objects, revealing that the abundance of the objects per site is independent of the size of the underlying network. The distribution of these objects show also power-law behavior, see Eq.~\ref{Eq:sigma2}. We identified critical exponents associated with the persistence of topological features across multiple scales, suggesting a deeper, multi-scale structure underlying the SOC behavior of the BTW model. These exponents, reported in Table~\ref{tab:exponents}, offer a novel way to quantify the complexity of the system, providing a bridge between traditional graph-theoretic measures and topological invariants.

In conclusion, our study demonstrates the power of combining visibility graphs with topological data analysis to investigate the dynamics of SOC systems at multiple scales. The new scaling behaviors and exponents we identified contribute to a more complete characterization of the self-organized criticality exhibited by the BTW model. Our findings highlight the importance of considering both local and global features when analyzing complex systems and suggest that visibility graphs and TDA could be valuable tools for studying other SOC models and dynamical systems.

Future work could explore the applicability of our approach to other types of SOC systems and time series, as well as investigate the relationship between the topological features of visibility graphs and the physical properties of the underlying system. By extending the analysis to higher-dimensional models or incorporating stochastic variations, we can further test the robustness of the scaling laws and critical exponents identified in this study.

\bibliography{refs}

\end{document}